\documentclass[ reprint,
 amsmath,amssymb,
 aps,
 prl,
floatfix,
]{revtex4-1}

\usepackage{graphicx}\usepackage{dcolumn}\usepackage{bm}\usepackage[colorlinks=true, citecolor=blue]{hyperref}

\begin{document}

\title{Anomalous nonlinearity of the magnonic edge mode}

\author{Mykola Dvornik}
\affiliation{Department of Physics, University of Gothenburg, 412 96 Gothenburg, Sweden}

\author{Johan \AA kerman}
\affiliation{Department of Physics, University of Gothenburg, 412 96 Gothenburg, Sweden}
\affiliation{Materials and Nanophysics, School of Engineering Sciences, KTH Royal Institute of Technology, 164 00 Kista, Sweden}

\date{\today}             
\begin{abstract}
Nonlinearity of magneto-dynamics is typically described by a single constant, $\mathcal{N}$, with positive and negative values indicating repulsion and attraction of magnons, respectively. In thin magnetic films with easy-plane magnetic anisotropy, magnon attraction is typically achieved for an in-plane magnetization. At sufficient stimulus, \emph{e.g.}~via application of spin transfer torque, the attraction can give rise to self-localized magnetic solitons, such as spin wave bullets, which shrink as their amplitude increases. In contrast, for an oblique magnetization above a certain critical angle, the repulsion of magnons only allows for propagating modes, which \emph{expand} when pumped more strongly. Here we demonstrate, both analytically and using micromagnetic simulations, that such a dichotomic description is inadequate for magnonic edge modes, which naturally appear in confined magnetic systems. In particular, we demonstrate that the confinement potential of such modes is nonlinear in nature and its contribution makes $\mathcal{N}$ non-monotonically dependent on their amplitude. As a prominent example, edge modes show compression and expansion for negative and positive $\mathcal{N}$, yet remain localized. In striking contrast to the extended geometries, edge magnons might also repeal even for an in-plane magnetization.

\end{abstract}

\pacs{Valid PACS appear here}                                                           \maketitle

Spin waves---collective oscillations of magnetic moments---hold great promise for the next generation of microwave technologies and emerging non-Von Neumann computing \cite{kruglyak2010jpd,chen2015pieee,Grollier2016procieee}. Their frequencies can span a very wide range: from a few GHz to tens of THz. In thin extended magnetic films the bottom limit is typically determined by the frequency of uniform ferromagnetic resonance (FMR).  Bellow this point lies a fundamental magnonic band gap where propagation of spin waves is not possible. 

Due to the presence of anisotropic magnetic interactions, magnetic systems are intrinsically nonlinear allowing for the spin waves to scatter with each other. Slavin and Kabos \cite{slavin2005ieeem} and later Gerhart et al. \cite{gerhart2007prb}~demonstrated that a single nonlinearity constant, $\mathcal{N}$, provides a satisfactory description of the nonlinear magneto-dynamics nucleated in thin magnetic films. The magnitude and sign of $\mathcal{N}$ determine the strength and type of spin wave scattering, with positive and negative values denoting repulsion and attraction of spin waves, respectively. In the former case, the spin wave frequency increases with amplitude resulting in a corresponding increase of the wavevector \cite{consolo2007prb}. Consequently, positively nonlinear spin waves will propagate farther, which can be viewed as an expansion of the magnetization dynamics due to the repulsion. If the nonlinearity is instead negative, the frequency of the magneto-dynamics drops with amplitude and might eventually be pushed into the magnonic band gap. As a result, self-localization of spin waves can be achieved, with the spin wave bullet \cite{slavin2005prl, consolo2008prb, bonetti2010prl, bonetti2012prb, demidov2012ntm} and droplet \cite{Ivanov1977, hoefer2010prb, mohseni2013sc,macia2014ntn,chung2016ntc,Chung2017arxiv} solitons being two prominent examples. Analogous to the case of positive $\mathcal{N}$, this opposite effect can be interpreted as a \emph{compression} of the magnetization dynamics due to the attraction between spin waves.

However, patterned magnetic films support linear localized modes in the fundamental magnonic band gap---so-called edge modes \cite{Jorzick2002}. They emerge in spin wave wells, which are relatively small partially demagnetized regions in the vicinity of the film's edges that are perpendicular to the saturation direction \cite{Joseph1965}. Recently, such modes were excited and driven into a strongly nonlinear regime by the spin-orbit torque in so-called spin Hall nano-oscillators (SHNOs) \cite{duan2014ntc, Dvornik2018prappl}. Yang et al.~suggested that, at least in transversely magnetized wires, the nonlinear properties of the edge modes should be similar to those of the spin wave bullet, \emph{i.e.} their volume and frequencies should decrease under pumping \cite{Yang2015srp}. However, a non-monotonic frequency behavior was systematically observed in constriction based SHNOs of various geometries, material compositions and applied magnetic field parameters \cite{Awad2017, Durrenfeld2017, Mazraati2016}. These results suggest that, in general, $\mathcal{N}$ of the edge mode is not constant, but is, on the contrary, amplitude dependent.

Guo et al.\cite{Guo2015prb} demonstrated that in magnetic nano-elements $\mathcal{N}$ of the bulk eigenmodes is mainly determined by the shape anisotropy. So, depending on the aspect ratio of the element, the nonlinear frequency shifts might have significantly different magnitudes and signs, compared to those of the extended films under the same applied field conditions. 

In this work, we demonstrate that the nonlinearity of the edge mode is given by an interplay of the dynamic and static dipolar fields. Similar to the extended geometries, the former contribution leads to the attraction of the spin waves, their frequency red-shift and shrinking vs.~mode amplitude. In contrast, the latter, edge-mode-specific contribution, describes the reduction of the confining demagnetizing field resulting in \emph{i}) de-localization of the mode, and \emph{ii}) its frequency increase.

\begin{figure}\centering
\includegraphics[width=8.6cm]{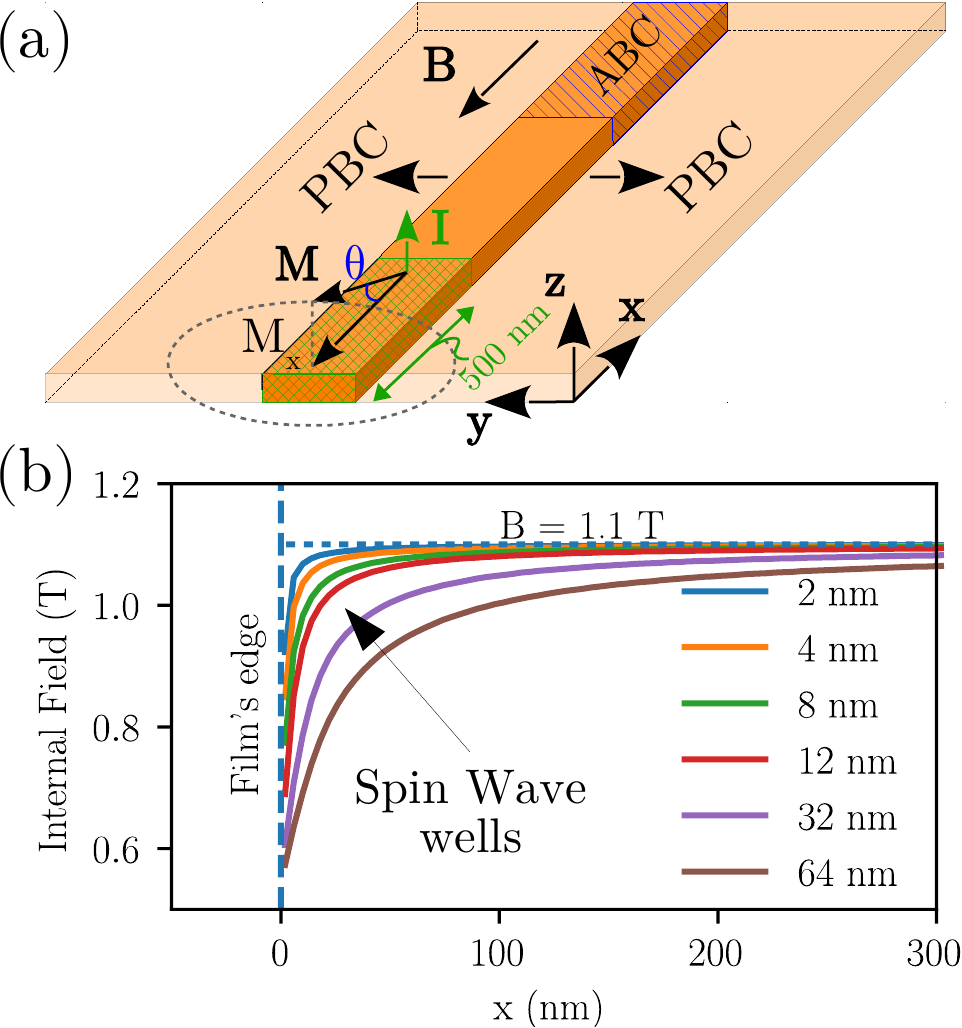} 
\caption{(a) The micromagnetic model of semi-infinite ferromagnetic film (light orange) approximated by a 4 $\mu$m $\times$ 32 nm $\times$ $t$ nanowire (dark orange), where $t$ is the thickness, with Periodic Boundary Conditons (PBCs) applied along its minor axis. The magnetization precession (shown by the dashed ellipse) is excited by the spin polarized current, $I$, applied uniformly to the hatched green region. Absorbing Boundary Conditions (ABCs) are used on the opposite edge of the film (blue hatched region) to damp-out spin waves. $\mathbf{M}$ and $M_x$ denote the magnetization vector and its projection on the equilibrium direction. (b) Internal magnetic field in the vicinity of the edge in ferromagnetic films of various thicknesses.}
\label{fig:1}
\end{figure}

The FMR frequency, $\omega_{0}$, of an ellipsoidal ferromagnetic system, without magneto-crystalline anisotropy and magnetized along one of its symmetry directions, is determined by the following, so-called Kittel equation:
\begin{equation}
    \label{Kittel}
    \begin{split}
    \omega_{0} = & \gamma \bigg[(B+(N_{\perp,\mathbf{y}} - N_{\parallel})\mu_{0} M_{x}) \times  \\ 
                 & \times  (B+(N_{\perp,\mathbf{z}}- N_{\parallel})\mu_{0} M_{x}) -N_{\perp,\mathbf{yz}}N_{\perp,\mathbf{zy}}M_{S}^2\bigg]^{\frac{1}{2}},
    \end{split}
\end{equation}
where $\gamma$ is the gyromagnetic ratio, $B$ is the strength of the applied magnetic field. $M_{x}$ is the lenght of the magnetization vector that in linear approximation is equal to the saturation magnetization, $M_{S}$; $N_{\perp,\mathbf{y}}$, $N_{\perp,\mathbf{z}}$ and $N_{\parallel}$ are the diagonal components of the demagnetizing tensor---two transverse and one parallel to the magnetization direction and $N_{\perp,\mathbf{yz}}$, $N_{\perp,\mathbf{zy}}$ are the corresponding off diagonal components. For an extended thin film magnetized in-plane and with $\mathbf{z}$ being its normal, $N_{\perp,\mathbf{y}} = N_{\parallel} = N_{\perp,\mathbf{yz}} = N_{\perp,\mathbf{zy}} \equiv 0$, leading to the commonly used expression:
\begin{equation}
    \label{FMR}
    \mathit{\omega_{FMR}} = \gamma \sqrt{B(B+N_{\perp} \mu_{0} M_{x})},
\end{equation}
where $N_{\perp} = N_{\perp,\mathbf{z}} = 1$.
It should be emphasized that here the magnetostatic contribution is due to the \emph{dynamic} dipolar field, $N_{\perp} \mu_{0} M_{S}$,  that appears when the magnetization vector goes out of plane during its precession cycle.

Now let us consider a semi-infinite film magnetized perpendicular to the corresponding edge, Fig.~\ref{fig:1}(a). In this case the longitudinal component of the demagnetizing tensor will become non-zero and spatially non-uniform in the vicinity of the edge giving rise to the \emph{static} demagnetizing field of $-N_{\parallel}(x) \mu_{0} M_{S}$\cite{Joseph1965}, Fig.~\ref{fig:1}(b). As demonstrated by Jorzick et al., the demagnetizing field creates a spin wave well that confines an edge mode \cite{Jorzick2002}. Its lowest order frequency is given by: \begin{equation}
    \label{EDGE}
    \begin{split}
    \mathit{\omega_{EDGE}} = &  \gamma \bigg[ (\tilde{B}-N_{\parallel}(x) \mu_{0} M_{x}) \times \\
                  & \times  (\tilde{B}-N_{\parallel}(x) \mu_{0} M_{x} + N_{\perp}(Qd)\mu_{0} M_{x})\bigg]^{\frac{1}{2}} ,
    \end{split}
\end{equation}
where $N_{\perp}(Qd)$ is the matrix element of the \emph{dynamic} dipole-dipole interaction, $d$ is the film thickness, and $Q$ is the imaginary wavevector of the edge mode determined by the Bohr-Sommerfeld quantization rule, $\tilde{B} = B + M_{x}\lambda_{ex}^2 Q^2$, where $\lambda_{ex}$ is the exchange length. Since edge modes are typically observed below FMR and taking into account that $0 \leq N_{\perp}(Qd) < 1$, one may conclude that the exchange contribution to the frequency of the edge mode is smaller than that of the \emph{static} demagnetizing field. So for the sake of simplicity, we will neglect it in the further discussion by assuming $\tilde{B} = B$. We also assume spatilly uniform components of the demagnetizing tensor, $N_{\parallel}(x) = N_{\parallel}$ which, in general, could be done using mean-field approach explained in Ref.\cite{Dvornik2018prappl}.

If the amplitude of precession is not vanishing, then \cite{A.G.Gurevich1996, Kosevich1990}: 
\begin{equation}
 \label{MPARALLEL}
 M_{x} = M_{S} \cos[\theta(x)] , 
\end{equation}
which describes the demagnetization of the ferromagnet due to the precession with angle $\theta(x)$; hence, the dependence of the precession frequency on its amplitude. So, the Eqs.~(\ref{FMR}) and (\ref{EDGE}) with substitution (\ref{MPARALLEL}) determine the frequencies of the nonlinear FMR and edge modes in thin magnetic films without magneto-crystalline anisotropies. 

\begin{figure}\centering
\includegraphics[width=8.6cm]{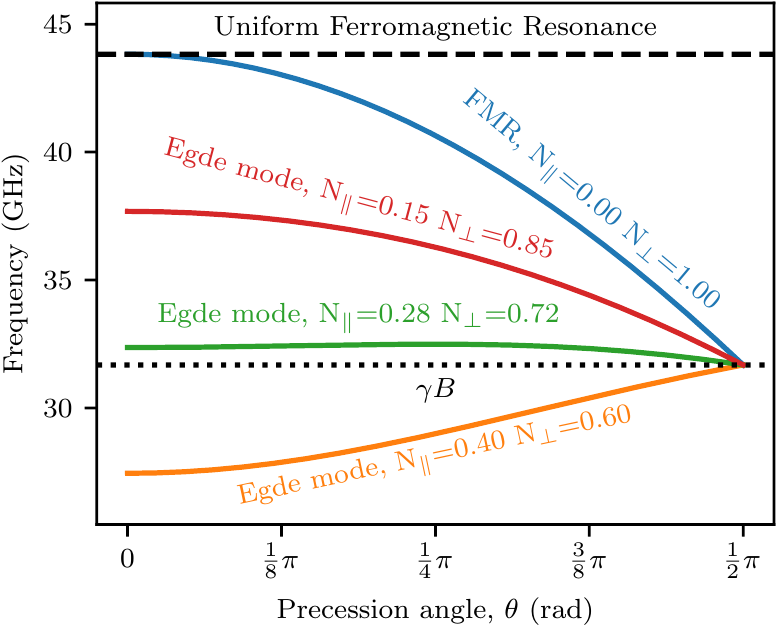} 
\caption{Frequencies of the FMR and edge modes with respect to their angles of precession estimated by substituting Eq. (\ref{MPARALLEL}) into Eq. (\ref{FMR}) and Eq. (\ref{EDGE}), respectively. In the calculations we assume, $B=1.1$ T, $\gamma$ = 28.8 GHz/T and $M_{s}$ = 8$\cdot 10^5$ A/m, which are typically for Permalloy.}
\label{fig:2}
\end{figure}

The demagnetization (reduction of $M_{x}$) lowers the contribution of the dynamic dipolar field, thus decreasing the frequency of the precession. In contrast, the shallowing of the spin wave wells increases the frequency of the localized mode. Since the FMR mode experiences only the dynamic dipolar field, its frequency always decreases with amplitude (i.e., it has negative nonlinearity) as depicted by the blue curve in Fig.~\ref{fig:2}. However, for the edge mode, the contribution of both static and dipolar fields should, therefore, lead to the nonmonotonic frequency vs. amplitude behavior as shown by the green curve in Fig.~\ref{fig:2}.

In fact, the non-monotonic frequency vs. amplitude behaviour of the edge mode is determined by the ratio of the demagnetizing tensor components. In particular, thicker films should show higher values of $N_{\parallel}$ and reduced $N_{\perp}$\cite{Joseph1965}. If $N_{\parallel} > \frac{1}{2}N_{\perp}$, then the mode will always show positive nonlinearity as demonstrated by the orange curve in Fig.~\ref{fig:2}. For $N_{\parallel} < \frac{1}{2}(\frac{B}{M_{S}} + N_{\perp} - \sqrt{\frac{B^2}{M_{S}^2}+N_{\perp}^2})$ only negative nonlinearity is possible as shown by the red curve in Fig.~\ref{fig:2}. For the values of $N_{\parallel}$ within these bounds the sign of the nonlinearity might change depending on the angle of precession as depicted by the green curve in Fig.~\ref{fig:2}. 

In the case of a weakly nonlinear magneto-dynamics, $\cos (\theta) \approx 1 - \theta^2 / 2 = 1 - \lvert c \rvert^2$, where $c$ is the complex dimensionless spin wave amplitude introduced in Ref. \cite{slavin2009ieeem}. The corresponding frequency shift is typically described by the nonlinearity coefficient,
$\mathcal{N}=\frac{\partial \omega}{\partial c^2}\rvert_{c^2=0}$,
where $\omega$ is the frequency of the spin wave.
As we already pointed out, for the FMR mode it originates only from the dynamic dipolar field. Thus, the nonlinearity of the FMR mode is negative
\begin{equation}
    \mathit{\mathcal{N}_{FMR}=-\frac{\gamma^2BN_{\perp}M_{S}}{\omega_{FMR}}\rvert_{c^2=0}}.
\end{equation}
In contrast, for the edge mode
\begin{equation}
    \mathit{\label{NEDGE}
    \begin{split}
    \mathcal{N}_{EDGE}= & -\frac{\gamma^2BN_{\perp}M_{S}}{\omega_{EDGE}} \\
              & +2\frac{\gamma^2 N_{\parallel} M_{S}}{\omega_{EDGE}}[B+M_{S}(N_{\perp}-N_{\parallel})(1-c^2)]\rvert_{c^2=0}.
    \end{split}}
\end{equation}
It should be noted, that the demagnetizing field should not exceed the applied one, i.e., $B\geq N_{\parallel}M_{S}$. Thus, the edge-mode-specific second term in Eq.~(\ref{NEDGE}) should be non-negative. So, the dependence of the spin wave well depth on the amplitude of the confined mode leads to the positive, amplitude-dependent contribution to the nonlinearity of the edge mode. According to Eq.~(\ref{NEDGE}) for the thin magnetic films ($N_{\perp} > N_{\parallel}$) it gradually decreases with the amplitude of precession. So for the films magnetized in-plane, the non-monotonic behaviour always manifest itself as gradual change from positive to negative nonlinearity as a consequence of \textit{dynamic} contribution taking over the \textit{static} one.

Interestingly, at the maximum amplitude of precession, $\theta=\pi/2$, frequencies of both FMR and edge modes converge to the same value (depicted by the dotted line in Fig.\ref{fig:2}). This behavior can be understood as a result of complete demagnetization that fully suppresses contributions of the dipolar fields. Thus, at this conditions the frequencies of the modes are only determined by the value of the applied magnetic field, $\mathit{\omega_{FMR}(\pi/2)=\omega_{EDGE}(\pi/2)=\gamma B}$, i.e., the modes exhibit a "paramagnetic-like" character.  

To validate our analytical considerations, we perform micromagnetic simulations of the current-induced magnetization auto-oscillations in in-plane magnetized extended Permalloy films using the \textsc{mumax$^3$} solver\cite{vansteenkiste2014aip}(see Fig.\ref{fig:1}a for the details of the micromagnetic model). We apply a uniform direct current to the edge region of the film (green hatched area in Fig.\ref{fig:1}a) and then estimate its threshold value required to start the auto-oscillations. Since we model a quasi-1D problem at zero temperature, the magnon-magnon scattering is significantly suppressed leading to the small value of the nonlinear damping. Consequently, the auto-oscillations reach large angles of precession even for small departures from the threshold current. So instead of looking at sustained dynamics, we trace its transient behavior over 125 ns as it approaches a steady state. We then employ a Short-Time Fourier Transform (windows size of 2.56 ns) with Kaiser window function (shape parameter $\alpha=3\pi$) to calculate the time evolution of the amplitude, frequency and spatial profile of the auto-oscillations. 

\begin{figure}\centering
\includegraphics[width=8.6cm]{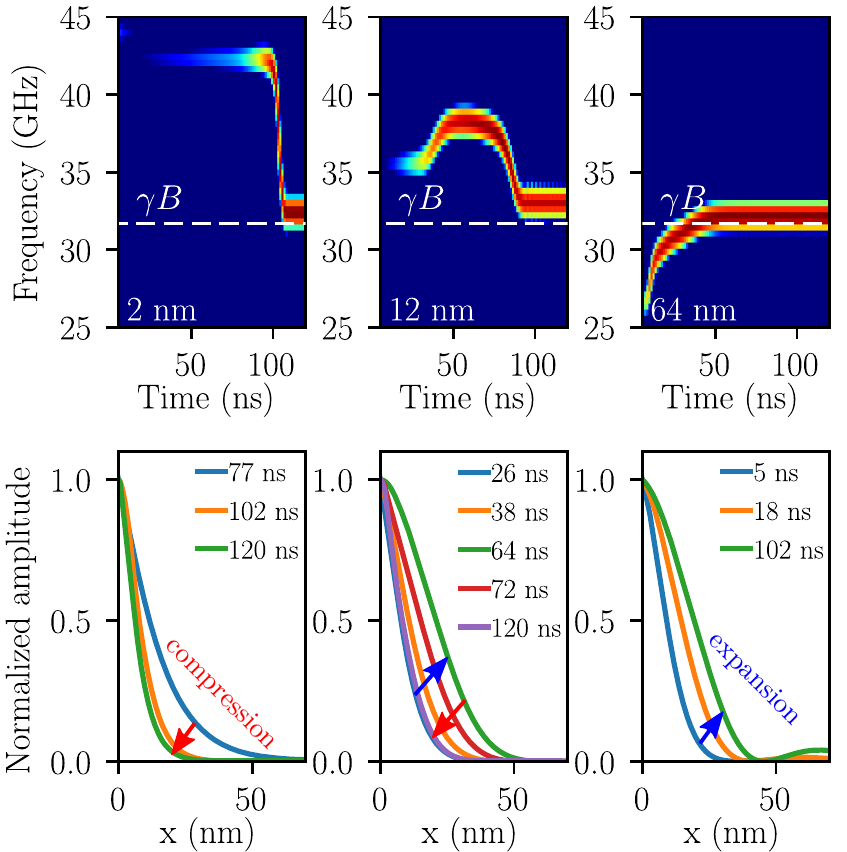} 
\caption{Frequencies of the FMR and edge modes with respect to their angles of precession estimated by substituting Eq. (\ref{MPARALLEL}) into Eq. (\ref{FMR}) and Eq. (\ref{EDGE}), respectively. In the calculations we assume, $B=1.1$ T, $\gamma$ = 28.8 GHz/T, $\lambda_{ex}$ = 5.69 nm and $M_{s}$ = 8$\cdot 10^5$ A/m, which are typically for Permalloy.}
\label{fig:3}
\end{figure}

The results of our modeling are shown in Fig. \ref{fig:3}. We simulate three film thicknesses of 8 nm, 12 nm and 64 nm each giving different values of the demagnetizing tensor components, i.e., thicker films have larger values of $N_{\parallel}$ and smaller values of $N_{\perp}$, leading to the deeper spin wave wells as shown in Fig.\ref{fig:1}b. This dependence is confirmed by the drop of the generation frequency with film thickness.

We observe an excellent qualitative agreement between the analytical model and the simulations. First of all, $\mathit{\mathcal{N}_{EDGE}}$ changes sign from negative to positive as the film thickness and, thus, $N_{\parallel}$ increase. Secondly, a non-monotonic $\mathit{\mathcal{N}_{EDGE}}(c)$ is observed at the intermediate film thickness of 12 nm highlighting the amplitude dependence of the nonlinearity coefficient. Finally, for all simulated film thicknesses, the frequencies of the auto-oscillations tend to $\omega_{B}$ confirming a vanishing $\mathcal{N}$ for precessional angles approaching $\theta=\pi/2$. It is worth mentioning that, although the peak amplitude of the auto-oscillating edge mode can reach $\theta=\pi/2$, the effective amplitude is typically smaller due to the strongly non-uniform precession. 

Our simulations reveal that the center of the edge mode (the location of its maximum amplitude) can detach from the edge and move inside the film with growing auto-oscillation amplitude. Similar behavior was experimentally observed by Demidov \emph{et al.} in in-plane magnetized nano-elements and was attributed to the nonlinear hybridization of the bulk and edge modes \cite{Demidov2010}. However, in our earlier work on auto-oscillations in constriction-based spin Hall nano-oscillators, we demonstrated that the linear edge modes move inside the bulk in obliquely applied fields, as a consequence of the shallowing of the spin wave wells \cite{Dvornik2018prappl}. As discussed above, the depth of the spin wave wells also decreases with the amplitude of the edge mode. So, the detaching of the edge mode is not related to a possible hybridization with the bulk mode. The details of this effect are however beyond the scope of the present work.

To understand how the sign of $\mathit{\mathcal{N}_{EDGE}}$ affects the spatial properties of the edge mode, we look at the corresponding mode profiles shown in the bottom row of Fig.~\ref{fig:3}. For the sake of comparison, each profile is normalized to the unit peak amplitude and aligned so that its maximum is located at $x=0$. Remarkably, we observe compression (shown by the read arrpows) and expansion (shown by the blue arrows) of the edge mode for negative and positive frequency shifts, consistent with the attraction and repulsion of magnons, respectively.

We believe that the nonlinear expansion and compression of the edge modes open a way to direct electrical control of the magnetic interaction in networks of nano-patterned magnetic structures, as such coupling is proportional to the effective volumes of the modes\cite{slavin2006prb}. In fact, it has been recently observed that mutual synchronization of constriction based SHNOs is typically achieved when edge mode experience expansion with drive current\cite{Awad2017}. Furthermore, linear edge modes are widely employed in magnonic crystals, which hold great potential for applications in next-generation signal communication technologies. Typically, frequency properties of such devices are tuned either at the fabrication stage or with impractically large applied magnetic fields\cite{klos2013magnonic, kumar2014magnonic, Tacchi2015, Bali2012high}. In contrast, driving the dynamics of magnonic crystals to strongly nonlinear regimes, would enable direct, and easy to implement, electrical control of their frequency properties.

In conclusion, we demonstrated that the nonlinear frequency shift of the magnonic edge modes depends on its amplitude and could be significantly tuned by nanopatterning. In striking contrast to the ferromagnetic resonance of the in-plane magnetized extended films, the magnonic edge mode can have a positive frequency shift and yet remain localized. Consistent with attraction and repulsion of magnons, edge modes exhibit spatial compression and expansion for negative and positive nonlinearities, respectively. Our findings pave the way for the all-electrical control of magnetic coupling in patterned networks of spin torque and spin Hall oscillators for the applications in nonlinear magnonics and neuromorphic computing.

The authors would like to acknowledge fruitful discussions with R. S. Khymyn.
\bibliography{references}

\end{document}